\documentclass[acmsmall,10pt,authorversion]{acmart}

\usepackage{graphicx}
\usepackage{listings}
\newcommand{\ignore}[1]{}
\newcommand{\desc}[1]{}

\AtBeginDocument{\catcode`\^^A=14 }

\newcommand*\numcircledmod[1]{\raisebox{.5pt}{\textcircled{\raisebox{-.9pt} {#1}}}}

\def \gaoa {{\sf Free Access}}
\def \freeaccess {{\sf Free Access}}
\def \aoa {{\sc AOA}}

\def \startsafe {{\em begin-write-only}}
\def \endsafe {{\em end-write-only}}

\def \dirty {{\sf dirty}}
\def \arbiter {{\sf ar}}

\makeatletter\if@ACM@journal\makeatother
\acmJournal{PACMPL}
\acmYear{2018}
\startPage{1}
\else\makeatother
\startPage{1}
\fi

\newtheorem{assumption}{Assumption}

\setcopyright{none}             

\begin{document}
	
	\title{Every Data Structure Deserves Lock-Free Memory Reclamation}

	
    \author{Nachshon Cohen}
\orcid{0000-0001-8302-2739}             
\affiliation{
	\institution{EPFL}            
	\city{Lausanne}
	\country{Switzerland}
}
\email{nachshonc@gmail.com}          

	\thanks{This paper appears in the Proceedings of the ACM on Programming Languages, Volume 2, Issue OOPSLA, 2018. When citing, please use that version}                

	\begin{abstract}
Memory-management support for lock-free data structures is well known to be a tough problem. Recent work has successfully reduced the overhead of such schemes. However, applying memory-management support to a data structure remains complex and, in many cases, requires redesigning the data structure. In this paper, we present the first lock-free memory-management scheme that is applicable to general (arbitrary) lock-free data structures and that can be applied automatically via a compiler plug-in. 
In addition to the simplicity of incorporating to data structures, this scheme provides low overhead and does not rely on the lock freedom of any OS services. 
	\end{abstract}

\ccsdesc[500]{Software and its engineering~Allocation / deallocation strategies}
\ccsdesc[500]{Software and its engineering~Concurrency control}
\ccsdesc[500]{Software and its engineering~Concurrent programming structures}
	

	\keywords{Lock-free memory management, Hazard Pointers, Optimistic Access}  

	\maketitle

\section{Introduction}
Lock-free data structures simplify reasoning about concurrent code. 
They are immune to dead locks, live locks, and they aid programmers in achieving high scalability and responsiveness. 
However, providing memory-reclamation support for such data structures, without foiling their progress guarantee, is difficult. 
Indeed, many lock-free data structures were published without any memory-reclamation support, 
as it was assumed that the application is short lived enough so that all memory is allocated in advance. 
This prevents the use of such data structures in real applications. 

Early solutions to this problem include the lock-free reference-counting scheme \cite{valois1995lock,mich95}, the epoch-based reclamation scheme \cite{harrisll} and the hazard-pointers scheme \cite{mich02,HP}. 
However, the epoch-based reclamation scheme foils the lock-freedom guarantee of the data structure; and the reference-counting and hazard-pointers schemes add significant overhead \cite{hart2007performance}. 
Recently, there has been a significant interest in designing lock-free reclamation schemes that support the lock-freedom progress guarantee, yet they do not add significant overhead \cite{dropta,Alistarh2015a,Cohen2015,Brown2015,Cohen2015c,Balmau2016,Dice2016a,Alistarh2017}. 

Although the above-mentioned solutions substantially reduce the overhead of lock-free memory-reclamation schemes, they address only specific (interesting) data structures. 
There is still a wide discrepancy between the expectations of the data-structure designer and the state-of-the-art memory-reclamation schemes. 
The pervasive assumption of data-structure designers is that memory reclamation is a non-issue: it is possible to publish a paper describing a novel design of a lock-free data structure without considering memory reclamation. 
It remains the responsibility of a (possibly non-expert) practitioner to apply a memory-reclamation scheme when incorporating the data structure into a real application. 
However, state-of-the-art memory reclamation schemes require the data structure to satisfy certain algorithmic properties that are not met by many practical data-structures. 
If a data structure does not satisfy the required properties, there is no clear recipe for modifying it so that it would be amenable to memory reclamation. 
In general, applying a lock-free memory-reclamation scheme to a given data structure might requires expertise that is not available to all practitioners, might modify the properties of the data structure, or might not be possible at all. 

Consider, for example, Harris's linked list \cite{harrisll} that stands at the core of many lock-free data structures. 
The original hazard pointers paper \cite{mich02} states that Harris's linked list is ``inherently incompatible'' with the hazard-pointers scheme\footnote{Section~\ref{subsec-hazardpointers} provides more background on this incompatibility}. 
Instead, they designed a variation of this data structure --- the Harris-Michael linked list \cite{magedll} --- on which hazard pointers can be applied. 
The differences between these data structures are subtle, hence it is easy to confuse them. 
Such confusion can lead even experienced researchers to apply hazard pointers on the wrong data structure, resulting in an incorrect design.  
Furthermore, it was later shown that the Harris-Herlihy-Shavit linked list \cite{herlihy2012art} is also incompatible with the hazard-pointers scheme. This data structure offers stronger progress guarantees (i.e., wait-free searches) and better performance than the Harris-Michael linked list \cite{Cohen2015c}. 
Other recent memory-reclamation schemes  \cite{dropta,Alistarh2015a,Cohen2015,Brown2015,Balmau2016,Dice2016a} are similarly incompatible with the Harris and Harris-Herlihy-Shavit linked-list data structures.  

In this paper, we present \freeaccess{}: the first memory-reclamation scheme that stands up to the expectations of the data-structure designers.
The \freeaccess{} scheme does not require the data structure to satisfy any algorithmic properties, hence making it generally applicable for existing and future lock-free data structures. 
Instead, it requires only that no instruction simultaneously reads a new pointer and writes to the shared memory.
All lock-free data structures we are aware of satisfy this property without any modification.
We cannot preclude a future novel design that uses an unsupported instruction, but even such a design could be modified easily by replacing the unsupported instruction with a Compare-And-Swap (CAS) loop, enabling \freeaccess{} to be applied\footnote{Section~\ref{section-assumption} discusses this unsupported instruction and presents the replacement algorithm.}. 
The main impact of the \freeaccess{} scheme is the decoupling of the design of lock-free data structures from the memory-reclamation effort; this decoupling enables designers of lock-free data structures to concentrate on designing the data structure without worrying whether their design is amenable to memory reclamation. 


The \freeaccess{} scheme is based on the concept of data-structure-specific garbage collection, introduced by \citet{Cohen2015c}. 
During the execution of the program, \freeaccess{} automatically determines the set of unreachable data-structure nodes and reclaims them for future allocations. 
The reclamation algorithm of \freeaccess{} is very similar to the Automatic Optimistic Access (\aoa{}) scheme \cite{Cohen2015c} and is loosely based on the mark\&sweep garbage-collection technique. 
It starts by gathering the {\em global roots}, data-structure nodes that are reachable by all threads (e.g., the head of a linked list); and the {\em local roots}, the data-structure nodes 
residing in the registers and stack of each thread. 
Then, it traces these roots and marks all data-structure nodes that are reachable from one of the roots. Finally, it reclaims all data-structures nodes that are unmarked. (More background information is provided in Section~\ref{sec-preliminaries}). 

Although the reclamation algorithm of \freeaccess{} is very similar to the \aoa{} scheme, there is a major difference between the two.
The \aoa{} scheme is applicable only for data structures written in a special normalized form of \citet{lf2wf}. 
As most standard data-structures are not written in this form, 
applying the \aoa{} scheme requires the practitioner to understand the data-structure algorithm well enough to re-write the data structure in this normalized form (or might not be possible at all). 
In contrast, the \freeaccess{} scheme does not require the data structure to be presented in any special format, which enables a (possibly non-expert) practitioner to easily apply \freeaccess{}, without understanding the internal data-structure details. 

Furthermore, we implemented an LLVM compiler pass that automatically modifies, according to the \freeaccess{} scheme, the data structure code. 
The programmer needs only to add the {\tt freeaccess} attribute to each function that corresponds to a data-structure operation and run the compiler pass.  
This stands in contrast to the existing lock-free memory-reclamation schemes that require the programmer to manually modify all loads (and possibly stores) in the data-structure code. 
Automatically applying the required modifications reduces the programming effort, keeps the code simple, and avoids bugs from imprecise modifications. 
We note that implementing a similar compiler pass for other lock-free reclamation schemes is complicated due to the need to inform the compiler about data-structure-specific mechanisms (e.g., how to restart an operation or how to correctly interpret the normalized form). 
This need to eliminated by the \freeaccess{} scheme.

The main challenge in designing an automatic lock-free reclamation scheme is collecting local roots in a lock-free manner. 
Although many garbage collectors attempt to provide partial guarantees of progress \cite{Herlihy1992lock, huds01,pizl07a,pizl08,auer08,pizl10}, 
they still require all threads to perform a synchronization barrier (e.g., handshake) for gathering local roots. 
If a thread fails to participate in the synchronization barrier, the local roots cannot be gathered and memory cannot be reclaimed. Eventually, threads will run out of memory and fail to make progress, thus foiling the lock-freedom guarantee \cite{petrank2012can}. 
The \aoa{} algorithm gathers local roots without any synchronization barrier,
but heavily relies on a special format of the data structure, which strictly specifies how local roots can be used. 

The \freeaccess{} scheme efficiently gathers local roots by partitioning the execution into read-only and write-only periods. 
Read-only periods can load new pointers but never modify the shared memory; write-only periods modify the shared memory but never load new pointers to a local variable. 
The \freeaccess{} scheme incurs a low overhead for write-only periods by noticing that the set of local roots is not modified during this period.  
It incurs an even lower overhead for read-only periods by designing a restart mechanism for arbitrary read-only periods. 
Using this restart mechanism, during read-only periods the thread is not required to publish the set of local roots at all; it only needs to restart if one of its local roots is reclaimed by concurrent threads. 

Our contributions in this paper are as follows.
\begin{itemize}
	\item A new algorithm for gathering local roots for general lock-free data structures, without depending on any algorithmic properties of the data structure. 
	\item The \freeaccess{} scheme, which combines our root gathering algorithm with the lock-free tracing algorithm of the automatic optimistic access scheme. The \freeaccess{} scheme can be applied to general lock-free data structure, thus enabling a separation of concerns between designing the data structure and applying it to a real application. 
	\item The implementation of an LLVM compiler pass that automatically applies the \freeaccess{} modifications, which reduces programming efforts and buggy implementations. 
\end{itemize}


\paragraph{Organization}
In Section~\ref{sec-preliminaries}, we provide preliminaries and discuss  some background. 
We present an overview of the \freeaccess{} scheme in Section~\ref{sec-overview} and details in Section~\ref{sec-algorithm}. 
In Section~\ref{sec-measurements} we present measurements of the algorithm.   
We discuss the assumption of separating code into read-only and write-only periods in Section~\ref{section-assumption}
and related work in Section~\ref{sec-related}. We conclude in Section~\ref{sec-conclusion}. 

\section{Background and Preliminaries}\label{sec-preliminaries}
\subsection{System Model and Progress Guarantees}\label{subsec-model}
\paragraph{Lock-freedom for memory reclamation}
We use the standard computation model of \citet{herlihy1991wait}.
A shared memory is accessible by all threads. 
The threads communicate through memory-access
instructions on the shared memory; a 
thread makes no assumptions about the status of any other threads, nor about the speed of its execution. 

A lock-free progress guarantee ensures that while the system executes a bounded number of steps, at least one thread progresses. 
Fair scheduling must not be assumed and the system must progress even if the scheduler is adversarial. 
Lock-free data structures are fast, scalable and widely used. 
They eliminate deadlock, livelock, and provide guaranteed system responsiveness. 
However, they permit a single thread to starve (even if it is scheduled enough steps) if other threads in the system progress. 
Wait-free progress guarantee ensures that a thread progresses after it executes a finite number of steps, regardless of the behavior of other threads in the system. 
A series of work \cite{Fatourou2011,Kogan2011,Kogan2012,Timnat2012}, culminating in Timnat and Petrank method \cite{lf2wf}, shown that wait-free algorithms can be designed efficiently.

Providing a lock-freedom guarantee for memory reclamation requires that under any schedules, including adversarial ones, 
at least one thread is able to progress in a finite number of steps. 
Waiting until all threads respond before reclaiming memory violates lock-freedom. 
Consider, for example, a schedule where a thread never executes any instructions. 
As this thread cannot respond, memory cannot be reclaimed. 
Eventually, other threads will not be able to allocate memory and will fail to progress, hence violating lock-freedom. 

We assume that the number of nodes available for allocation (i.e., available memory) is sufficiently larger than the maximum number of nodes that are simultaneously used by the data structure. 
Otherwise, a data structure that allocates memory, but never frees, would violate lock-freedom in a non-interesting manner. 

\paragraph{Read reclaimed nodes}
The \freeaccess{} scheme assumes that the memory allocator is strictly lock-free and that it does not return memory to the OS, which results in lock-acquisition for the current Linux implementation. 
Consequently, all allocated memory resides in the application's address space and accessing it does not generate a segmentation fault, {\em even if the accessed node was already reclaimed}. 
Alternatively, it assumes that a segmentation fault can be caught by a signal handler and ignored.  
Specifically, we follow the \aoa{} scheme and make the following assumption. 

\begin{assumption}\label{subsection-read-free}
	If a thread reads the memory of a reclaimed node, the read value is arbitrary. However, the thread is allowed to proceed to the next step of execution.
\end{assumption}


\subsection{Lock-Free Memory-Reclamation Schemes}
In this subsection, we provide some necessary background regarding lock-free memory-reclamation schemes. 
We begin by briefly presenting the hazard-pointers scheme and describe why it does not support the Harris-Herlihy-Shavit linked-list data structure. 
The \freeaccess{} scheme extends the \aoa{} \cite{Cohen2015c} scheme and inherits some of its properties. 
In this subsection, we sketch the design of the \aoa{} scheme and present some parts of the algorithm that \freeaccess{} inherits. 
We also sketch the normalized-form representation, on which \aoa{} depends. 
Recall that the main difference between \freeaccess{} and \aoa{} is that \aoa{} requires the data structure to be presented in a special normalized form, whereas \freeaccess{} does not pose this restriction. 

\subsubsection{The Hazard-Pointers Scheme}\label{subsec-hazardpointers}
The hazard-pointers scheme is considered the most popular memory-reclamation scheme for lock-free data structures. 
The hazard-pointers scheme associates a set of {\em hazard pointers} to each thread; these pointers are used by the thread to announce that a set of data-structure nodes are currently accessed by the thread and must not be reclaimed. 
A hazard pointer can be written only by the owner thread, but can be read by other threads in the system. 
The hazard-pointers scheme modifies the data-structure code by replacing all reads of a new data-structure node with the code of Listing~\ref{listing-hpread}. 
Reclaiming a node requires three steps. 
First, the node must be disconnected from the data structure. 
Second, the application passes the node to the {\em retire} method, which announces the node as a candidate for reclamation. 
Third, the scheme waits until no hazard pointer points to this node then reclaims the node (make it reusable).
Note that a candidate for reclamation (Step 2) is not immediately reclaimed as a thread could access it concurrently. 
\begin{lstlisting}[label=listing-hpread,caption={Reading a pointer in the hazard-pointers scheme},float]
Node *HPRead(Node **addr, int i){
	while(true){
		Node *newnode = *addr; //read address of the new node
		thread.HP[i].store(newnode, memory_order_seq_cst); //ensure that newnode is visible. Expensive
		if(*addr == newnode) //validate that new node is still reachable from previous node
			return newnode; 
		else continue; //failed to ensure that newnode is protected. Restart. 
	}
}
\end{lstlisting}^^A \label{listing-hpread}

\begin{figure}
	\includegraphics[width=0.9\textwidth]{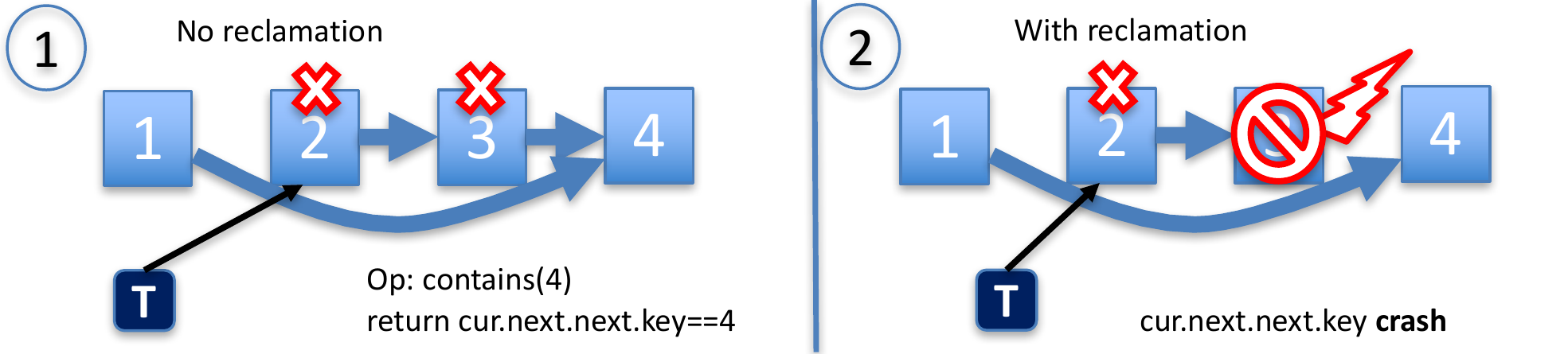}
	\caption{Illustrating why the hazard-pointers scheme cannot be applied directly to Harris-Herlihy-Shavit linked list. The black arrows represent $T$'s current pointer and hazard pointer, protecting node \boxed{2}.
		Nodes \boxed{2} and \boxed{3} were removed from the list. 
		$T$ continue traversal from \boxed{2} until finding \boxed{4}. 
		If the hazard-pointers scheme is used, this optimization cannot be used since \boxed{3} is not protected by a hazard pointer and can be reclaimed. }
	\label{figure-hp-bad}
\end{figure}

The hazard-pointers scheme cannot be applied to all data structures and specifically cannot be applied to Harris's linked list or Harris-Herlihy-Shavit linked list. 
To explain why, consider Harris-Herlihy-Shavit linked list and suppose that no memory reclamation is used\footnote{Harris's linked list is discussed in Appendix~\ref{appendix-hp-bad}. Harris-Herlihy-Shavit linked list was designed for Java, which provides correct memory reclamation without progress guarantees.}. 
Consider a snippet of the linked list with 4 nodes $\boxed{1} \Rightarrow \boxed{2} \Rightarrow \boxed{3} \Rightarrow \boxed{4}$ and a thread $T$ that searches whether key 4 exists, while currently traversing node \boxed{2}. 
While $T$ is asleep, concurrent threads remove \boxed{2} and \boxed{3} and disconnect them from the list. 
When $T$ wakes up, a possible way to continue the traversal is to restart from the head pointer, then traverse \boxed{1} and \boxed{4}.
There exists, however, a better way.
Even though \boxed{2} is not part of the linked list, the target node of $T$, \boxed{4}, is still reachable from \boxed{2}. 
Thus, $T$ can continue navigation (ignoring the fact that \boxed{2} and \boxed{3} were removed) until it reaches the target node \boxed{4}. See the illustration in Figure~\ref{figure-hp-bad} (left). 
If \boxed{4} is marked as deleted (or simply does not exists in the search path), the argument become subtle as there might be another node \numcircledmod{4} that is not marked as removed and is reachable from \boxed{1} (but not from \boxed{2}). 
Nonetheless, it is possible to show that returning a ``key does not exist'' is correct since there exists a time during the execution of the operation when the key did not exist in the list. 

\def \hpBadWidth {0.4\textwidth}

However, this mechanism does not work when the hazard-pointer scheme (or a similar scheme) is applied. 
When \boxed{2} and \boxed{3} are removed from the list, both become candidates for reclamation. 
As $T$ traverses \boxed{2}, there exists a hazard pointer guarding \boxed{2} and it will not be reclaimed. 
But $T$ does not traverse \boxed{3} when it is removed, hence \boxed{3} is reclaimed; after re-allocation, \boxed{3}'s content becomes arbitrary. 
When $T$ continues its traversal to find \boxed{4}, it reads \boxed{3}'s next pointer that returns an arbitrary address (instead of \boxed{4}). 
An attempt to read from the arbitrary address would crash the application.

The Harris-Michael algorithm does not use the wait-free traversal optimization (nor does it use the single-CAS-multiple-removes optimization used by the Harris's algorithm, see Appendix~\ref{appendix-hp-bad}). 
When the Harris-Michael algorithm encounters a removed node, it immediately attempts to disconnect the node from the list. 
If disconnecting fails, the operation is {\em restarted}. 
Thus, in the example above, $T$ must restart its traversal from the head pointer. 
Nevertheless, this variant is compatible with the hazard-pointers scheme. 

\ignore{We emphasize that while \freeaccess{} may provide some performance improvement by supporting such an optimization, the main focus is simplicity. 
A programmer porting Harris-Herlihy-Shavit algorithm to C/C++ would need to (a) understand that hazard pointers cannot be applied, (b) disable the wait-free search optimization, which is embodied deep in the algorithm's design, (c) prove the correctness of the unoptimized version, and (d) prove that hazard pointers can be applied to the unoptimized version. }

We note that restarting an operation does  {\em not} generally mean restarting the operation from scratch. For example, if the thread already applied the operation to the data structure, restarting the operation from scratch can lead to re-application of the same operation twice, resulting in an incorrect behavior of the data structure.

\subsubsection{Normalized Form of Lock-Free Data Structures}\label{subsec-normalized}
The normalized form of \citet{lf2wf} was originally designed for upgrading lock-free data structures into wait-free ones. 
This form is used by the \aoa{} scheme for formally defines restarts for lock-free data structures. 
If a data structure is presented in a normalized form, each operation consists of three parts: generator, executor, and wrap-up. Loosely speaking, the executor is responsible for applying the operation and must not be applied twice, and the generator and wrap-up do not modify the abstract representation of the data structure and can be restarted an arbitrary number of times\footnote{Disconnecting a marked node is an example of a step that modifies the shared memory without changing the abstract representation of the data structure.}. 
The executor has a very rigid structure. It executes a list of CASes (produced by the generator); stops on the first CAS failure or completion of all CASes; and returns the number of successfully executed CASes. 
The generator and wrap-up are permitted to modify the shared memory (assuming this modification can be applied multiple times without harming correctness) only via the CAS instruction. 
In addition, the execution of generator and wrap-up must only depend on the operation's input.
The wrap-up part can additionally depend on the CAS list executed by the executor and on the number of successfully executed CASes (i.e., input and output of the executor). 

The normalized form requires that all modifications to the shared memory be executed by the CAS instruction. 
In recent years, there has been a growing trend of using the Fetch-And-Add (FAA) instruction for synchronization, in addition to CAS \cite{Braginsky2016,Basin2017,Yang2016}. 
Under a high contention, the CAS instruction is likely to ``fail'', hence necessitating additional attempts to apply the operation and increasing the contention on the memory bus. 
In contrast, the  FAA instruction always ``succeeds'' in providing each thread a unique token, which reduces contention. 
Data structures using FAA cannot be represented natively in the normalized form.

\subsubsection{Automatic Optimistic-Access Scheme}\label{subsection-aoa}
The \aoa{} scheme was designed to be easily applicable and to support Harris and Harris-Herlihy-Shavit data structures. 
At runtime, it automatically identifies the nodes that are unreachable and reclaims them, 
without imposing any kind of synchronization barrier for initiating or finishing reclamation phases. 
Thus, the \aoa{} scheme reclaims memory, even in the presence of a thread that does not respond during multiple reclamation phases. 

The \aoa{} scheme consists of four parts: programmer interface, initiating mechanism, gathering local roots, and tracing (which are discussed next). The \freeaccess{} scheme inherits the programmer interface, the initiating mechanism and the tracing algorithm of the \aoa{} scheme; and the reader should consider these background parts as necessary for this paper. 
In contrast, the \freeaccess{} scheme gathers local roots in a completely different manner, which is described in the rest of the paper. 
The reader can consider the \aoa{} method mainly for the sake of contrasting it with this work. 

\paragraph{Programmer Interface} The programmer has to specify the class or struct used in representing data-structure nodes. 
Specifically, the programmer specifies the size of each node and the offset of pointer fields that point to other data-structure nodes (pointers to other objects are ignored during reclamation). 
At runtime, the program specifies the size of the memory pool; 
it also specifies the location of the {\em global roots} to the data structure that is visible to all threads (e.g., the head pointer in a linked list). 
Finally, all memory allocations must use a special AOAalloc function. 
Note that no free-equivalent interface is provided: all memory is reclaimed automatically by the scheme, similarly to GC languages. 
The \freeaccess{} scheme uses exactly the same interface as \aoa{}.

\paragraph{Initiating Mechanism}
A memory reclamation {\em phase} is initiated when the memory available for allocation is exhausted. 
Any thread in the system can initiate a new phase and many threads can simultaneously attempt to initiate a new phase. 
A phase  is similar to the concept of a GC cycle. 
But there is a difference: To satisfy lock-freedom, no synchronization barrier is executed at the beginning of a new phase, hence some threads might lag behind and still operate on an older phase.

A new phase starts by signaling all threads.
It is desirable to avoid signaling each thread multiple times, even if multiple threads attempt to initiate a new phase simultaneously. 
Toward this end, each phase has a unique, always increasing, {\em index}. 
Each thread has a 8-bit \dirty{} flag that co-resides with a 56-bit index on a single machine word. 
Signaling a new reclamation phase is done by setting the \dirty{} flags of all threads. 
This is done exactly once for each thread by using a CAS instruction that sets \dirty{} while increasing the thread's index to the current index.  
A thread can efficiently check whether a new phase begun by simply checking whether its \dirty{} flag is set. 
A pseudo code is provided in Listing~\ref{listing-set-dirty}. Recall that this mechanism is also used by \freeaccess{}. 

\begin{lstlisting}[caption={Start reclamation phase},label={listing-set-dirty}]
union udirty{
	atomic<bool> dirty; //fast access to the dirty flag. 
	atomic<uint64_t> dirty_phase; //8-bits dirty, 56-bits phase index
	pair<bool dirty,uint64_t phase> get()
		long val = dirty_phase.load(); 
		return make_pair(val&1, (val>>8)); 
	bool CAS(pair &ex, pair de)
		return dirty_phase.compare_exchange_strong(ex, de, memory_order_seq_cst); @\label{line-set-dirty}@
}; 
atomic<uint64_t> phaseIndex; 
thread_local uint64_t lPhaseIndex; 
void InitReclamation(){
	//advance shared phase#; one thread will succeed. 
	phaseIndex.compare_exchange_strong(lPhaseIndex, lPhaseIndex+1); 
	lPhaseIndex = phaseIndex; //current thread might lag, just read most updated number. 
	foreach(thread T) //Signal all threads by setting the dirty flag
		val = T.udirty.get(); //read dirty flag and the associated phase index. 
		while(val.phase < lPhaseIndex) //dirty was not set yet in current phase
			T.udirty.CAS(val, {lPhaseIndex, true}); //set dirty, advance to current phase. 
}
\end{lstlisting}^^A \label{listing-set-dirty} \label{line-set-dirty}

\paragraph{Gathering Local Roots}
To gather local roots in a lock-free manner, the \aoa{} scheme requires data-structure operations to be presented in the normalized form of \citet{lf2wf}. 
Recall from Section~\ref{subsec-normalized} that each such operation consists of three parts: generator, executor, and wrap-up. 
Loosely speaking, the \aoa{} scheme considers local roots only for the executor part of each operation. 
The generator and wrap-up can be restarted an arbitrary number of times, hence the thread's local state is ``unimportant''. 
Before the executor begins, all local pointers are stored in the hazard pointers of the thread. 
The executor and the wrap-up parts must only depend on nodes in the CAS list (as generated by the generator part) and some additional non-pointer fields. 
Thus, specifically these nodes are stored in the hazard pointers of the thread. 
Then, the thread checks, via the \dirty{} flag, whether a new reclamation phase has started. 
If \dirty{} is set (i.e., a new phase has started), the local state might be unsafe: it is abandoned and the operation restarts from the beginning of the generator. 
Recall that according to the normalized form structure, the generator can be restarted an arbitrary number of times without affecting the correctness of the data structure. 
If \dirty{} is clear (i.e., no new phase has started), the thread proceeds to the executor. 

The generator and wrap-up parts are considered to have no local roots, hence a memory reclamation phase might reclaim nodes accessed by these parts. 
Thus, every load from memory is followed immediately by checking the \dirty{} flag. If the \dirty{} flag is set, a new phase has started and the local state might contain pointers to reclaimed nodes. 
The thread immediately abandons its local state and restarts the current part (generator or wrap-up). 
Correctness results from the normalized-form property that permits the generator or wrap-up to restart an arbitrary number of times. 

The hazard pointers, which are assigned at the beginning of the executor, are kept until the wrap-up is completed. 
The reason is that the wrap-up might depend on the input to the executor that contains only the nodes protected via hazard pointers. 
To correctly restart the wrap-up part, these nodes must not be reclaimed while the wrap-up is being executed. 

\paragraph{Tracing and Reclamation}
Once local roots (hazard pointers of all threads) and global roots (as reported by the programmer interface) are gathered, the \aoa{} algorithm traces these roots and marks all nodes that are reachable from the roots. The \aoa{} algorithm finishes by reclaiming all unmarked nodes for future allocations. 
The tracing and reclamation algorithm of the \aoa{} scheme is based on a mark\&sweep GC with a side markbit table. 
But it also handles many subtleties that arise when attempting to implement mark\&sweep in a lock-free manner. 
For example, the algorithm must protect against a thread that wakes up after a reclamation phase is finished and that attempts to mark a stale node. 
The \aoa{} algorithm solves this issue by using a versioning scheme for the side markbit table. 
Another issue that the \aoa{} algorithm deals with is enabling tracing to continue even if a thread is stuck while it processes a node. 
Solving this issue requires that the currently processed node is always visible to other threads and a helping mechanism for helping the stuck thread. 
For the sake of brevity, we do not list here the full details of the \aoa{} algorithm. 
The \freeaccess{} scheme modifies only the way local roots (i.e., hazard pointers) are set and gathered; it uses exactly the same tracing algorithm as \aoa{}, thus avoiding the need to re-consider the above subtleties in our context. A high-level pseudo code for the reclamation algorithm appears in Listing~\ref{listing-reclamation}. 

\begin{lstlisting}[caption={Reclamation},label={listing-reclamation}]
void Reclamation(){
	InitReclamation(); //Listing @\ref{listing-set-dirty}@ above
	Roots = gatherLocalRoots(); //For freeaccess, Listing @\ref{listing-gathering-local}@ below
	Roots += gatherGlobalRoots(); //e.g., head of linked list. As reported by the programmer. 
	trace(Roots); //mark nodes reachable from LocalRoots
	sweep(); //reclaim unmarked nodes
}
\end{lstlisting}^^A \label{listing-reclamation}

\section{Overview}\label{sec-overview}
The main challenge that \freeaccess{} addresses is gathering local pointers of all threads in a lock-free manner. 
According to the definition of lock freedom (Section~\ref{subsec-model}), 
once the memory reclamation-scheme begins reclamation, it must gather the local roots of all threads while the system executes a finite number of steps. 
This problem is difficult. 
Gathering local roots seems to follow two paths: cooperative (asking each thread to report its local roots) or uncooperative (without getting any help from the thread). 
Collecting local roots in a cooperative manner is problematic because it foils the lock-freedom guarantee. 
Under the lock-freedom model, it is not possible to assume that the thread would execute any instructions before the system executes a finite number of steps (or even before reclamation finishes). 
But collecting local roots in an uncooperative manner is also problematic as current architectures do not enable a thread to access 
registers of another thread. 
Another seemingly possible solution is to prevent the thread from using local roots: all local pointers must always be visible in the shared memory. 
But such a solution is also very problematic as the thread needs to update the shared copy (and issue an expensive memory fence) anytime a new pointer is loaded, which results in an extremely high overhead.

\gaoa{} solves this problem by partitioning the execution into write-only and read-only periods. 
For write-only periods, the assumption is that while the thread is busy writing, it does not load new pointers into its local variables. 
Hence, it is sufficient to publish all local pointers at the beginning of the write-only period. 
These local pointers are valid throughout the execution of the entire write-only period, thus reducing the publishing overhead.

\begin{figure}[t]
	\fbox{\includegraphics[width=0.98\textwidth]{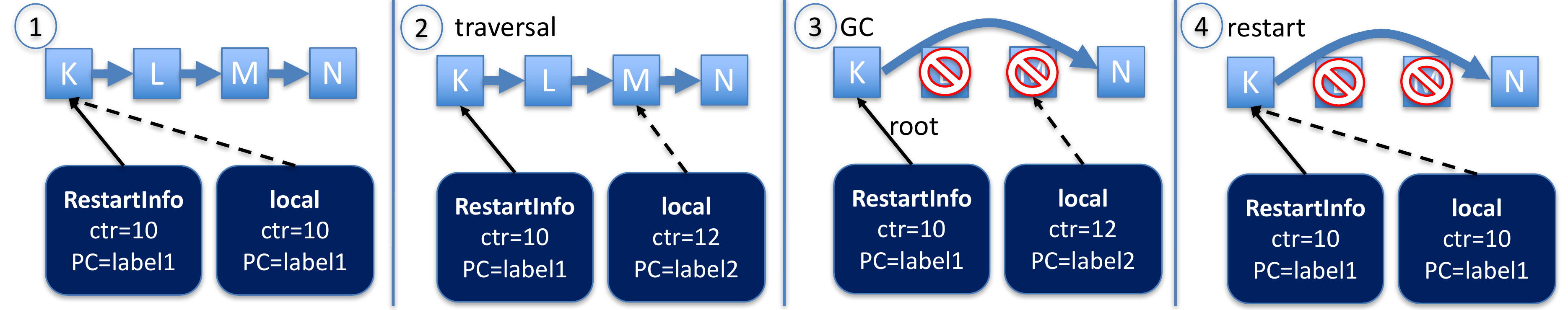}}
	\caption{Illustrating read-only periods. On \numcircledmod{1}, the thread is immediately after a write-only period, hence the restart information matches the local variables. PC represents the program counter and ctr some local integer variable. 
		On \numcircledmod{2}, the thread traversed two nodes in the list.
		The restart information is unmodified. 
		On \numcircledmod{3}, concurrent threads delete L and M and invoke a GC. 
		The GC gathers K as a root from the restart information ({\tt RestartInfo}); M is thread-local and inaccessible to the GC. It reclaims L and M as they are unreachable from any root. 
		On \numcircledmod{4}, the thread restarts using the restart information ({\tt RestartInfo}). 
		The thread state is valid: there are no pointers to reclaimed nodes. 
		The performance benefits are due to the infrequent changes to the root, which reduces the number of fences. 
	}
	\label{figure-illustration-read-only}
\end{figure}

For read-only periods, the main observation is that such periods can be restarted in a local manner because they never modify the shared state of the program. 
Returning to the beginning of the current read-only period requires only loading the local state of the thread at the beginning of the read-only period, without requiring cooperation from other threads. 
Hence, it is sufficient to consider the state of the thread --- and the set of local pointers it uses --- only at the beginning of the read-only period. 
An illustration of this idea appears on Figure~\ref{figure-illustration-read-only}.

Furthermore, the beginning of a read-only period is actually the end of the previous write-only period. 
The \freeaccess{} scheme publishes the set of local roots at the beginning of the previous write-only period, and we assume that these local roots are not 
modified during the write-only period. Thus, the set of local roots at the beginning of the read-only period is already visible and no additional publication step is required. 
Overall, the \gaoa{} scheme never requires the thread to publish its state during read-only periods, thus allowing the system to achieve good performance.

Finally, consider the problem raised at the beginning of the section regarding the hardness of gathering local pointers of a thread. 
The \gaoa{} scheme solves this problem for read-only periods by using two properties. 
First, it ensures that the local pointers of the thread {\em at the beginning of the period} are visible in the shared memory and that they can be gathered. 
Second, it ensures that the thread restarts its execution from the beginning of the period. 
These two properties enable threads to use local state without frequent publishing while also enabling the memory reclamation scheme to gather roots in a lock-free manner. 


\section{The Free Access Algorithm}\label{sec-algorithm}
In this section, we provide the details of the \gaoa{} scheme for gathering the set of local roots of all threads in a finite system-wide steps. 
We begin by defining the data structures used by each thread in the system. 

\subsection{Thread Local Variables}
Each thread has a {\em dirty flag}, two {\em frames of hazard pointers}, an {\em arbiter flag}, and a non-pointer {\em checkpoint}. 
The {\em dirty flag} is inherited from the \aoa{} scheme, as mentioned in Section~\ref{subsection-aoa}. 
A dirty flag signifies, when it is set by other threads, that a new reclamation phase started and nodes might be reclaimed. 
The dirty flag is read and cleared by the thread and is set by other threads (Listing~\ref{listing-set-dirty}). 

A hazard-pointers frame is written only by the thread and can be read by other threads. 
It contains all local pointers during a write-only period. 
During a read-only period, the hazard-pointer frame is used to store information about the local pointers at the beginning of the read-only period. 
If a restart is required, the local pointers are restored from the hazard-pointer frame. 
Each thread has an {\em arbiter flag} (\arbiter{}), that is used to decide which of the two hazard-pointer frames is currently used. 
Two frames of hazard pointers are used for enabling the thread to publish the local roots in a new hazard pointers frame without foiling the (still important) information stored in the current hazard pointers frame. 
The arbiter flag is local to the thread and cannot be accessed by other threads. 
During a read-only period, the current hazard pointers frame is denoted HP[\arbiter] and the other frame is denoted HP[!\arbiter] (pronounced as {\em not} \arbiter). 

The {\em checkpoint} stores all local information required for returning to the beginning of the current read-only period, including all local (non-pointers) variables and the program counter. 
However, it does not contain local pointers, because these are restored from the current hazard pointers frame.

\subsection{Write-Only Periods}
Beginning a write-only period (and ending the previous read-only period) starts with publishing the set of local pointers of the thread in HP[!\arbiter]. 
This includes writing all local pointers and issuing a memory fence to ensure that these pointers are visible to other threads. 
Then, the thread checks the \dirty{} flag. If the \dirty{} flag is off, then the thread flips the meaning of the arbiter flag and proceeds to the write-only period. 
But If the \dirty{} flag is on, a new reclamation phase is started and might have already gathered the local roots of the thread.
Recall that gathering local state is done by reading the hazard pointers of the thread; and these pointers correspond to the state of the thread at the beginning of the previous read-only period. 
Thus, the thread restarts its execution at the beginning of the previous read-only block. 
To restart the previous read-only block, the thread replaces the content of its local variables with the content stored in the checkpoint (for non-pointers) and the current hazard-pointer frame (for pointers); and it jumps to the 
program counter stored at the checkpoint.

When a write-only period ends, the thread stores all of its non-pointer local variables in the {\em checkpoint}. 
It does not store the local pointers as they are already stored on HP[\arbiter] (with the new arbiter value).
Note that a write-only period does not read pointers from the heap, hence the set of local pointers at the beginning of the write-only period are the same as the local pointers at the end of the write-only period. 
Therefore, even though the current hazard-pointer frame was written at the beginning of the write-only period, it does not need to be updated. 

The pseudo code for the restart mechanism is provided in Listing~\ref{listing-restart},
and the pseudo code for \startsafe{} and \endsafe{} is in Listing~\ref{listing-startend-safe}. 
\begin{lstlisting}[caption={Restart}, label={listing-restart}]
void restart(){
	helpReclamation(); //help if there is reclamation phase running, as in @\cite{Cohen2015c}@
	dirty.store(false, memory_order_relaxed); //no need to restart again, we are already restarting
	for each $p$ in local pointers do:
		p=HP[ar].p; 
	for each $v$ in local variables do:
		v=checkpoint.v;
	goto checkpoint.program_counter; 
}

\end{lstlisting}
\begin{lstlisting}[caption={\startsafe{} and \endsafe{}},label={listing-startend-safe}]
void beginWriteOnly(){ //=endReadOnly
	For each $p$ in local pointers do: 
		HP[!ar].p=p; 
	atomic_thread_fence(memory_order_seq_cst); //read dirty after setting HPs
	if(dirty.load(memory_order_acquire))
		restart();
	ar=!ar;
}
void endWriteOnly(){ //=beginReadOnly
	for each $v$ in local variables do:
		checkpoint.v=v; 
	checkpoint.program_counter=&PC; //PC is the program counter of the next instruction 
}

\end{lstlisting}^^A \label{listing-restart} \label{listing-startend-safe}

\subsection{Read-Only Periods}
If a reclamation phase starts when a thread executes a read-only period and the thread does not respond, 
the \freeaccess{} scheme gathers the set of roots published in the hazard-pointer frame of the thread. 
This represents the set of roots at the beginning of the read-only period of the unresponsive thread. 
To ensure that the behavior of the memory-reclamation algorithm is correct, the unresponsive thread must restart at the beginning of the read-only period. 
Furthermore, it must not execute any instructions that are visible to other threads before the restart. 
Therefore, the system behaves as if the thread never started the read-only period for the entire duration that its roots were gathered; 
in which case, it is clearly correct to gather the set of roots at the beginning of the read-only period. 
An illustration of this mechanism is provided in Figure~\ref{figure-illustration-read-only}. 

Ensuring that the thread is not visible until it is restarted requires that the thread does not write to the shared memory; this follows immediately from the definition of 
read-only periods. 
However, it is also required that the thread behaves ``reasonably'' so that it does not cause segmentation faults or other runtime errors that are visible, even though 
the thread does not explicitly write to the shared memory \cite{dicepitfalls}. 
As memory reclamation does not wait for the thread to respond and does not gather the set of current local pointers of the thread, 
it is possible that a currently accessed node is reclaimed. 
In general, this can lead to an arbitrary behavior of the thread and cause the thread to be visible before it manages to restart. 

A core invariant maintained by the \freeaccess{} scheme is that the content of a reclaimed node is never used. 
To understand how a reader thread $T$ maintains this invariant, we first consider an abstract view of how reclamation works. 
Reclamation starts by setting $T$'s \dirty{} flag to true (Listing~\ref{listing-set-dirty} Line~\ref{line-set-dirty}) and then reclaims some nodes and sets their content to arbitrary. 
To ensure that $T$ never uses an arbitrary value, every read by $T$ is followed by reading the \dirty{} flag. Only if the \dirty{} flag is clear, is the read value used. 
Otherwise, $T$ restarts the operation, using the algorithm in Listing~\ref{listing-restart}. 
As reclamation sets the content to arbitrary only after setting $T$'s \dirty{} flag, and as $T$ uses the read value only if the \dirty{} flag is off after reading the content, then $T$ never uses the arbitrary content. 

There are two ordering constraints that a reader thread must maintain. First, it must read the \dirty{} flag after reading the node content. Second, it must use the read node content after checking that the \dirty{} flag is clear. 
Ensuring that the compiler and the CPU do not reorder the reads requires the use of fences. 
Specifically, the first ordering constraint is satisfied by using an acquire fence after reading the node content and before reading the \dirty{} flag. Correctness is similar to \cite{Boehm2012}. 
The second ordering constraint is satisfied by using an additional acquire fence when reading the \dirty{} flag. 
The validation code appears in Listing~\ref{listing-example-read} (the macro VALIDATE\_READ). 

A simple way to maintain the read invariant is to execute the validation code immediately after every read (thus, before using the read value). 
But in some cases, a node has multiple fields that are read simultaneously and then used. 
In such a case, the validation code can be placed once, after reading both fields and before using both values. 
The invariant is still maintained as these two reads are independent and as each read is validated before the read value is used. 
An example code for applying read validation also appears in Listing~\ref{listing-example-read}.

\ignore{
Checking the \dirty{} flag also require careful memory ordering annotations. 
Recall that reclamation may run concurrently with a reader. 
Such a concurrent reclamation starts by setting the \dirty{} flag using memory\_order\_seq\_cst (Listing~\ref{listing-set-dirty}), which is later followed by reclaiming some nodes and setting their content to arbitrary. 
The reader code is designed to prevent using such arbitrary values. 
Thus, reading the \dirty{} flag must happens {\em after} reading from a node and {\em before} the read value is used. 
Memory ordering fences must be used to prevent reordering of the reads by the compiler or the CPU. 
To ensure the first part, we use a memory\_order\_acquire fence between reading the value and reading the \dirty{} flag. 
Correctness follows by a similar argument to \cite{Boehm2012}. 
The second part (i.e., read value is used only after checking the warning flag), follows since no load can be reordered before a load with memory\_order\_acquire. 
An example of applying checks to reads appears at Listing~\ref{listing-example-read}; Line~\ref{line-fence-p} protects the read of $p$ and Line~\ref{line-fence-q} protects the read of $p$ and $q$, which are independent. }

\begin{lstlisting}[caption={Applying \gaoa{} to read-only period},label={listing-example-read}]
#define VALIDATE_READ 		   atomic_thread_fence(memory_order_acquire); \
									if(dirty.load(memory_order_acquire)) restart();
	q=p.load(memory_order_relaxed);
	VALIDATE_READ; //validate q is not an arbitrary value @\label{line-fence-p}@
	v1 = q.load(memory_order_relaxed); //use read value, also another load from memory. 
	v2 = p.load(memory_order_relaxed);   //another independent load
	VALIDATE_READ; //validate v1 and v2 are not arbitrary values @\label{line-fence-q}@
	//use v1, v2. 
\end{lstlisting}^^A \label{listing-example-read}

\subsection{Starting and Ending a Data-Structure Operation}
At the beginning of a data-structure operation, the thread's local state is initialized as follows. 
The arbiter flag is set to zero. 
To avoid publishing an uninitialized pointer, local pointers are initialized to NULL . 
If a data-structure operation starts with a read-only period, as is common in many algorithms, a dummy write-only period is executed at the beginning of the function to ensure that 
the read-only period can be restarted. 
In this case, the checkpoint of the thread consists of the input to the function and of the program counter at the beginning of the function. 
If the input to the function does not contain any pointer to a data-structure node,  there is no need to write any hazard pointer and the memory fence can be elided. 

Finishing an operation should nullify all hazard-pointer fields to avoid {\em floating garbage}: memory that cannot be freed even though it is not used by any data-structure operation. 
However, the \freeaccess{} scheme uses two sets of hazard pointers, hence nullifying all of them is inefficient. 
Instead, the thread writes a special marker (-1) to the first hazard pointer in HP[1]. 
This marker signifies that all other pointers are implicitly NULL. 
When the next operation writes to the hazard pointers, it alway starts with the first hazard pointer at HP[1] (recall that the arbiter flag is initially zero and is swapped before setting hazard pointers), thus removing the implicit marker. 

\subsection{Gathering Local Roots}
When a reclamation phase starts, the \freeaccess{} scheme gathers, by reading the hazard pointers of all threads, their local roots. 
There are two subtleties to be considered. 
First, if the first hazard pointer in HP[1] of a thread contains the special marker (-1), 
all other hazard pointers of this thread are ignored. 
Second, if the data structure uses tagged pointers (i.e., modifies some bits of pointer fields), these tagged pointers are untagged when gathering the local roots. 

We remark that the \aoa{} scheme converts all tagged pointers into untagged pointers before setting them into the thread's hazard pointers. 
Thus, gathering the local roots is trivial in the \aoa{} scheme. 
This is not possible in the \freeaccess{} case, because hazard pointers are used for restarts.
If a pointer is tagged when executing a read-only period and the read-only period requires a restart, the pointer must remain tagged. 
Hence, the pointer is stored to the hazard-pointer frame in its tagged form and is untagged during local-root gathering. 
In fact, clearing the tag while gathering the roots provides a minor performance benefit as gathering roots occurs less frequently than setting hazard pointers does. The pseudo code for root gathering is provided in Listing~\ref{listing-gathering-local}. 

\begin{lstlisting}[caption={Gathering local roots},label={listing-gathering-local}]
set<node*> gatherLocalRoots(){
	set<node *> Roots; 
	foreach( thread T )
		if(T.HP[1].first == MARKER) continue; //continue with next thread
		for( ar @$\in$@ {0,1}, $p$ in local pointers )
			Roots += clearTagging(T.HP[ar].p); 
	return Roots; 
}
\end{lstlisting}^^A \label{listing-gathering-local}

\subsection{Automatically Applying Free Access Modifications}\label{subsec-llvm}
Although the modifications mentioned above do not require special algorithmic properties, implementing them manually is a tedious effort, 
risks bugs, and complicates the original data-structure code. 
To simplify this process, we implement an LLVM compiler pass that automatically applies all these modifications. 
The programmer needs only to add the {\tt freeaccess} attribute to each data-structure operation. 
To avoid modifications to the clang front-end, we use the general annotate attribute, that enables the programmer to specify an arbitrary string as an attribute. 
Thus, the programmer simply adds \_\_attribute((annotate("freeaccess"))) to each data-structure operation.

The compiler pass starts by parsing the llvm-ir annotation header and by extracting the set of functions with the {\tt freeaccess} attribute. 
Then, for each of these functions, it executes the following four steps. 
First, the compiler pass extracts the set of local variables. 
For each non-pointer variable, it adds an additional local variable that serves as the checkpoint value for original variable. 
It also adds the arbiter flag and the program-counter (used for restarting) as two additional local variables and inserts code to initialized them. 
Lastly, Initialization sets all local pointers to NULL. 
Before returning from the operation, the compiler pass inserts a code to set the special marker (-1) to the first entry in HP[1], which implicitly nullifies all hazard pointers. 

Second, the compiler pass creates a new basic block for restarting the operation, denoted {\em restart basic block}. 
The code of the restart basic block closely follows the code in Listing~\ref{listing-restart}. 

Third, the compiler pass considers reads from the shared memory. 
It begins by analyzing consecutive reads to find if they are independent. If they are, the first one is discarded, because checking the \dirty{} flag after the second one is sufficient to protect the first one. 
Then, for each remaining read, the compiler pass inserts a check to the \dirty{} flag and a conditional branch to the restart basic block. 

Fourth, the compiler pass handles writes (and CASes) to the shared memory. 
It analyzes the code and identifies write-only regions.
Then, it inserts code for starting a write-only region (\startsafe) at the beginning of the region, and it inserts code for ending a write-only region (\endsafe) at the end of the regions. 
The inserted code closely follows Listing~\ref{listing-startend-safe}. 

The current version of the plug-in implements the hazard-pointers set as an array of 15 elements, which forms 2 cache lines, together with the \dirty{} flag. 
This supports up to 7 pointers for each data-structure operation. 
We carefully coded the plug-in to ensure that if the pointer to the \dirty{} flag and the arbiter are stored in registers (as in the code produced by the assembler we used), 
then checking the \dirty{} flag or accessing a hazard pointer is done in a single x86 instruction\footnote{We slightly changed the meaning of the \arbiter{} flag in the code: instead of being either 0 or 1, it is either 1 or 9,
	so the address of the $i$-th hazard pointer in the current frame (HP[\arbiter]) is  {\tt ((node **)\dirty{})+\arbiter+i}. Flipping \arbiter{} is done by XORing it with 8.}. 

If an operation is read only (as determined by LLVM's {\tt functionattrs} pass), our compiler pass produces an optimized version. 
As there are no writes, restart is done by jumping to the beginning of the operation, using a non-conditional jump instead of an indirect jump. 
Hence, there is no need for the restart program-counter. 
Furthermore, hazard pointers are never accessed, so the arbiter flag is also omitted. 

Our LLVM compiler pass is scheduled to run before other optimization passes, enabling it to produce unoptimized code that is later heavily optimized by other compiler passes. 
Except for the two optimizations mentioned above, the implementation of the compiler pass is quite straightforward and it consists of less than 600 lines of C++ code.

\section{Measurements}\label{sec-measurements}
\def \fa {FA}
\def \rc {RC}
\def \hp {HP}
\def \hpmb {HPMB}
\def \nr {NR}
\ignore{Thus far, the paper presented the benefits of the \freeaccess{} scheme in terms of simplicity. 
However, the overhead our scheme imposes is also important and it is unlikely to be used if the overhead is much higher than other reclamation schemes. 
The \freeaccess{} scheme is compared against several other schemes. 
First, we would like to know the raw overhead it incurs. Thus, \freeaccess{} is compared against a leaky implementation that incurs no memory reclamation overhead. 
Second, the \freeaccess{} algorithm is based on the AOA algorithm, while being significantly easier to apply. We would like to know what is the cost of this simplicity. That is, what is the overhead that \freeaccess{} incurs in addition to the AOA overhead. 
Third, the only existing lock-free reclamation scheme that can be considered automatic is lock-free reference counting \cite{valois1995lock,mich95}.}

We incorporated the \gaoa{} scheme with the \aoa{} implementation\footnote{Code was taken from github.com/nachshonc/AutomaticOptimisticAccess}. 
We measured the effectiveness of our system using the linked-list implementation by \citet{herlihy2012art}
and a hash-map implementation based on this list. 
The data-structure code was modified to replace all AOA modifications with the {\tt freeaccess} attribute and run the \freeaccess{} compiler pass. 
The resulting implementation is denoted \fa.

The \fa{} implementation is compared against several other reclamation schemes. 
The baseline for comparison is a leaky implementation that never reclaims memory and incurs no memory reclamation overhead; it is denoted \nr. 
Typically, the leaky implementation cannot be used for real applications. 
The comparison of \fa{} and \nr{} represents the full cost of adding \freeaccess{} to a data structure that did not previously support memory reclamation. \nr{} allocates all memory before execution begins so that memory allocation from the OS does not become the bottleneck. 
The second comparison is against the AOA scheme. 
As \freeaccess{} is based on the AOA scheme but provides further simplicity, a comparison of \fa{} and \aoa{} reveals the direct cost of the additional simplicity. 
We also consider the lock-free reference-counting scheme \cite{valois1995lock,mich95}, which is also applicable for 
general code but suffers from high overhead; it is denoted \rc. 
These are the only schemes we are aware of that guarantee strict lock freedom and support the Harris-Herlihy-Shavit linked list \cite{herlihy2012art}. 

For completeness, we also compare \fa{} to the hazard-pointers scheme; the latter is applied to the Harris-Michael linked list \cite{magedll} (Section~\ref{subsec-hazardpointers} discusses the main difference from the Harris-Herlihy-Shavit algorithm); it is denoted \hp. 
Finally, we also include a version of hazard pointers that uses fence elision \cite{Dice2016a} that replaces frequent memory fences during data-structure operations by the {\tt membarrier} Linux system call during reclamation.
As the {\tt membarrier} system call cannot return until it interacts with all other threads in the system, we do not consider it to be strictly lock-free. 
Still, it guarantees lock-freedom, except for the {\tt membarrier} syscall, and it is considered a reasonable tradeoff in practice.
Similarly to the \hp{} implementation, this version is applied to the Harris-Michael linked list because it cannot be applied to the Harris-Herlihy-Shavit linked list; it is denoted \hpmb.

\subsection{Methodology}
For the linked-list data structure, we consider two configurations, LL5K and LL128; they differ in the number of keys in the list. 
In the LL5K configuration, the range is set to 10K so that the average size of the list is 5K. This configuration frequently reads from the shared memory and exhibits low contention. 
In the LL128 configuration, the range is set to 256 so that the average size of the list is 128. 
The contention is significantly higher and frequency of memory reclamation increases. 
For the hash set, the average table-size is 10K and the load factor is 1. 
The operations are extremely fast, hence the frequency of memory reclamation increases further. However, contention is relatively low as different threads operate on different buckets. 

We ran the experiments on a machine featuring 4 AMD Opteron (TM) 6376 2.3GHz processors, each with 16 threads (overall 64 hardware threads). 
We varied the number of threads between 1 and 64, in power-of-two jumps to check the behavior of the proposed schemes under different scalability settings. 
The operation distribution was fixed to 50\% contains and 50\% modifications, equally distributed between inserts and removes. 
This models a write-heavy distribution that stresses memory reclamation. 
Each execution runs for 1 second and each test was repeated 10 times. We depicted the ratio between the average throughput and the throughput of the \nr{} implementation and the 95\% confidence intervals. 
The code was compiled using LLVM version 5.0.0 with the -O3 optimization flag, running on an Ubuntu 16.04 OS (kernel version 4.4.0-72). 

For the \gaoa{} and the AOA schemes, before the measurements began we initialized the allocation pool with 50,000 nodes ; 
a new collection phase is triggered when the allocation pool 
is exhausted.
For \hp{} and \hpmb{} implementations, reclamation is triggered locally by a thread once its local list of retired nodes (recall Section~\ref{subsec-hazardpointers}) reaches a threshold; this is different than with  the \fa{} and AOA schemes, where 
reclamation is triggered globally. 
To ensure a fair comparison, we set the local reclamation threshold to 100,000 nodes divided by the number of threads, which approximates a slightly lower reclamation rate than the \fa{} scheme\footnote{Setting reclamation to 50,000 nodes divided by the number of threads ensures similar maximum memory size but a higher average reclamation rate. We verified that a lower reclamation rate improves performance.}. 

For the \hpmb{} implementation, the {\tt membarrier} syscall is invoked with the {\tt MEMBARRIER\_CMD\_\-PRIVATE\_EXPEDITED} parameter that was recently introduced to Linux. 
This parameter provides a significant performance boost to \hpmb{} compared to the older variant of the {\tt membarrier} syscall.

\subsection{Results}
\newcommand{\mysize}{55mm}
\newcommand{\figureHeight}{7cm}
\newcommand{\figureWidth}{0.85\textwidth}

\begin{figure}
\centering
\includegraphics[width=\figureWidth]{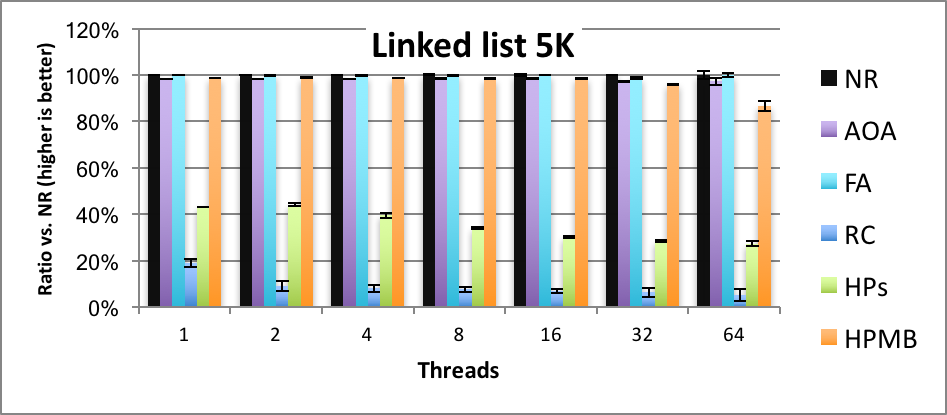}
\caption{Comparing the \nr{}, AOA, \fa{}, \rc{}, \hp{}, and \hpmb{} memory reclamation schemes for a linked list with 5,000 elements. 
	For each scheme, we depict the throughput ratio between the throughput of the depicted scheme and the leaky implementation (\nr).  }
\label{figure:ll5}
\end{figure}
\begin{figure}
	\centering
	\includegraphics[width=\figureWidth]{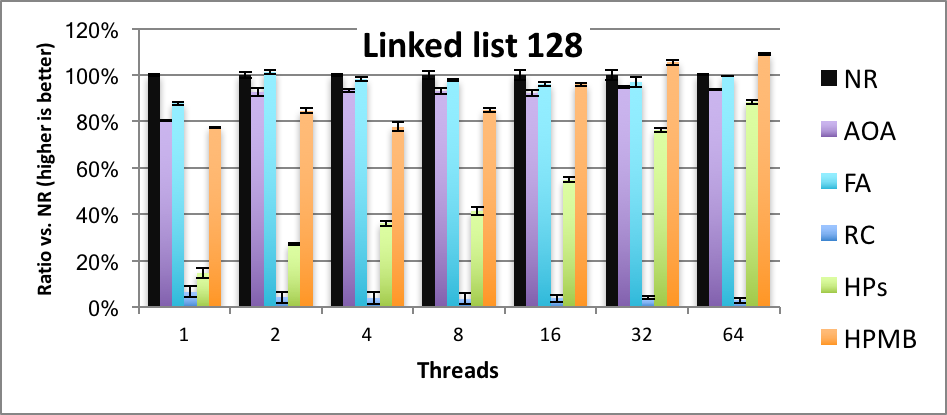}
	\caption{Comparing the \nr{}, AOA, \fa{}, \rc{}, \hp{}, and \hpmb{} schemes for a linked list with 128 elements. }
	\label{figure:ll128}
\end{figure}

\newcommand{\rpm}{\raisebox{.2ex}{$\scriptstyle\pm$}}
The results for the linked list with 5,000 elements are presented in Figure~\ref{figure:ll5}. 
Each operation is relatively long (around 20$\mu$s) and most of the time is spent on reads. 
Compared to the leaky implementation, \fa{} has an overhead of 0.1\% -- 1.4\%. 
Compared to the AOA scheme, \fa{} is faster by 1.3\% -- 2.5\%. 
We attribute the difference to implementing the \freeaccess{} scheme at the LLVM IR level (Section~\ref{subsec-llvm}) that generates better code than the manually modified version. 
In this case, simpler is strictly better. 
\fa{} is significantly faster than \rc{}: over 5x for a single thread and more than 10x for 2 threads and above. 
It is well known that the overhead of \rc{} is extremely high; 
still, this was the only available memory-reclamation scheme that could be applied to any data structure (without garbage cycles).
The \freeaccess{} scheme is clearly superior to \rc{}: it can be applied automatically (including garbage cycles), easily (using our compiler plug-in), and provides significantly better performance. 
Due to this huge overhead, we ignore \rc{} throughout the rest of this section.

The \hp{} implementation performs slower than \fa{} by 56\% -- 72\%. This is mainly the cost of the memory fences required during navigation. 
Finally, the \hpmb{} is slightly slower than \fa{} by around 1\% up to 16 threads. 
For 32 threads and above, the overhead of \hpmb{} increases to 3\% -- 13\%. 
The main reason for this overhead is the baseline data structure: \hpmb{} is applied to Harris-Michael linked list data structure, which disables the wait-free navigation optimization discussed in Section~\ref{subsec-hazardpointers}. 
Note that without this optimization, contain operations are more heavyweight as they need to synchronize with removal operations and might need to restart.

The results for the linked list with 128 are presented in Figure~\ref{figure:ll128}. 
The operations are faster, but contention is significantly higher: 
a single operation takes around 230ns for the single threaded case and around 7.5$\mu$s for the 64-thread case. 
The overhead of the \fa{} scheme is 12\% for the single threaded case and less than 4\% for 2 threads and above. 
Recall that memory reclamation efforts are mostly thread-local. Consequently, when contention on the list grows, the overhead of these local operations decreases. 
\fa{} is faster than the AOA scheme by 2\% -- 7\%. 
The \hp{} implementation is slower than \fa{} by 6x for the single threaded case, but this overhead reduces when contention on the list increases and reaches only 11\% for 64 threads. 
Comparing \fa{} and \hpmb{}, we see that \fa{} performs better for 1 -- 8 threads by 3\% -- 16\%, but slower when the number of threads is 16 or higher. 
Interestingly, for 32 and 64 threads, \hpmb{} performs even faster than \nr{} by up to 8\%. 
Profiling shows that part of this benefit is due to quickly reusing reclaimed nodes, while they are still in the processor cache. 
The reader should recall that \hpmb{} calls the {\tt membarrier} method, which waits until all threads execute a memory fence. As a result, \hpmb{} does not guarantee lock-freedom progress. 
If the programmer is willing to compromise for the progress guarantee, 
\freeaccess{} trades off some performance for the sake of simplicity.

\ignore{
For the Intel machine, the \freeaccess{} scheme actually performs faster than \nr{}. 
This machine has 8GB of memory and the \nr{} implementation allocates over 2GB at the beginning of the execution. This seems to stress the kernel and cause performance reduction. We verified that 
allocating the same amount of memory for the \freeaccess{} scheme reduces its performance to below \nr{}, even if this memory was not used during the execution. 
\fa{} is still faster than AOA by 13\% -- 20\%. 
}

\begin{figure}
	\centering
	\includegraphics[width=\figureWidth]{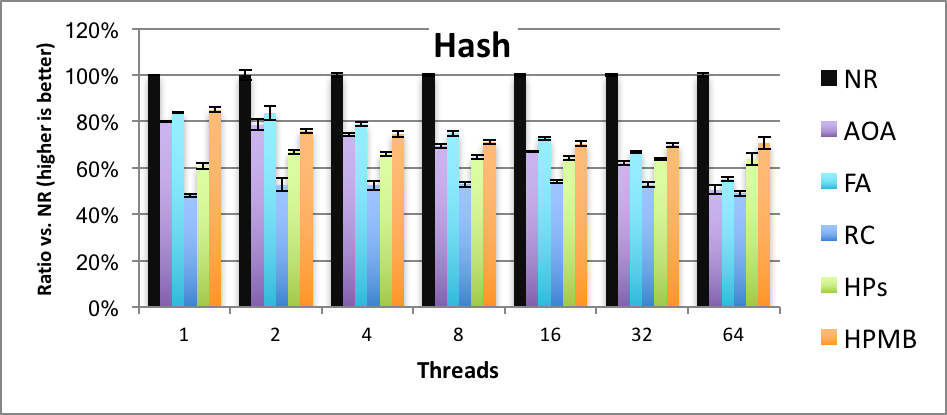}
	\caption{Comparing the \nr{}, AOA, \fa{}, \rc{}, \hp{}, and \hpmb{} schemes for a hash table with 10,000 elements. }
	\label{figure:hash}
\end{figure}

The results for the hash table are presented in Figure~\ref{figure:hash}. 
Hash operations are extremely short: on average, an operation visits only a single node; still, contention is low as each operation uses a different bucket. 
The overhead of the \fa{} scheme (compared to \nr{}) is 16\% -- 44\% and increases for a high-thread count. 
Profiling shows that the main overhead comes from the tracing algorithm; because operations are extremely short, the overhead of tracing cannot be hidden. 
Furthermore, tracing is not as scalable as hash-map operations, which explains the increased overhead when the number of threads increases. 
Comparing \fa{} to AOA, we find the \fa{} scheme is faster by 3\% -- 6\%. 
Most overhead comes from the tracing algorithm that is shared by both implementations. 
The overhead of the \hp{} implementation is higher than \fa{} by 22\% for a single-threaded case, similar to \fa{} for 32 threads, and lower than \fa{} by 9\% for 64 threads. 
The higher throughput at the 64-threaded case again shows the benefits of manually reclaiming memory, also for the strictly lock-free version of the hazard-pointers scheme. 
There is no need to trace the data structure at runtime to gather reclaimed nodes; and reclamation decisions can be done in a local manner. 
By avoiding frequent memory fences, the \hpmb{} version provides further performance improvements over \hp{} . 
Up to 32 threads, \fa{} performs similar to \hpmb{}: between 8\% faster and 3\% slower. 
But for 64 threads, \hpmb{} outperforms \fa{} by 15\%. 
Again, the comparison against the hazard-pointers scheme represents a tradeoff. 
The \freeaccess{} scheme is the simplest to apply and provides good performance for a low-thread count; but when operations are extremely short it suffers from reduced scalability. 
The \hp{} implementation is hard to apply, but provides some performance benefits for 64 threads. 
The \hpmb{} implementation compromises the lock-freedom guarantee, in addition to being hard to apply; but it provides best performance for 64 threads. 
%

\ignore{
For the Intel machine, results are significantly better for all memory reclamation schemes. 
The overhead of \fa{} is only 7\% -- 11\%. 
\fa{} is still faster than AOA by 0\% -- 8\%. 
The overhead of \hp{} is only 13\% -- 19\%, but it is always slower than \fa; of course, our Intel machine has only 16 hardware threads, so \hp{} may still performs better on a larger machine. 
\hpmb{} still performs slow due to the overhead of frequent invocations of {\tt membarrier}.
}

Overall, the overhead of simplicity is quite low. The \freeaccess{} scheme consistently outperforms the AOA scheme, even though \fa{} is significantly simpler to apply and does not require the data structure to be presented in a normalized form. 
Compared to the hazard-pointers scheme, the \freeaccess{} scheme incurs some overhead for the 64 threads case. 
In this case, the \freeaccess{} scheme makes a tradeoff similar to garbage collected systems: 
it pays some cost at runtime, for the sake of simplicity for the programmer. 
In other cases, where the number of threads is lower or operations are longer, the \freeaccess{} scheme provides better performance by avoiding expensive memory fences and supporting the 
faster Harris-Herlihy-Shavit data structure.  
Finally, \hpmb{} compromises the lock-freedom guarantee by avoiding frequent memory-fences. 
For a low-thread count, it significantly outperforms \hp{}, but \fa{} still provides better performance. 
For the 64-threads case, \hpmb{} outperforms \fa{} by a wider margin (15\% for the hash configuration) but remains hard to apply as \hp{}. 

\section{Assumption: Separating Write-Only and Read-Only Periods}\label{section-assumption}\label{subsection-write-only-or-read-only}
A central assumption of the \freeaccess{} algorithm is the possibility to partition the execution of a program into write-only and read-only periods. 
Read-only periods must not at all modify the shared memory.
But write-only periods restrict only the reads of new pointers that have a special meaning for root gathering. 
Write-only periods are permitted to read any non-pointer field from the shared memory. 
Furthermore, many lock-free data structures use the CAS instruction that writes a new value to the shared memory, but only if the current value matches
the expected value. 
Typically, these data structures use the boolean version of CAS, which returns a boolean indicator whether the expected value matches and the new value is written\footnote{Java only supports the boolean version of CAS, which can be implemented using LoadLinked/StoreConditional primitives.}. 
This instruction is still considered write-only as it does not load a new pointer to the local state of the thread (hence can be ignored by root gathering). 
All the lock-free data structures we are aware of in the literature modify the shared memory, either by a direct store, the boolean version of CAS, or by fetch-and-add (FAA) on a integer field. Thus, the \freeaccess{} scheme is applicable to all these cases.

Current architectures, however, also support instructions that can modify the shared state and simultaneously load a pointer. 
Consider the SWAP instruction (x86 {\tt xchg} instruction) that swaps the content of a memory location and a register. 
If SWAP is applied to an integer field, it is considered as a write-only instruction. 
But if it is applied to a pointer field, then this instruction both writes a memory location and reads a pointer to a register. This usage cannot be partitioned into either write-only or read-only period and is not supported by the \freeaccess{} scheme. 
This instruction cannot be restarted because the SWAP can be visible twice to other threads. 
But if the thread is stuck immediately after executing the SWAP, it is not possible to discover the read pointer that resides in a register that is unaccessible by other threads. 

Although the lack of support for SWAP is a limitation of the \freeaccess{} scheme, it seems to be a general limitation for lock-free data structures in general. 
The problem is similar: If a lock-free data structure uses a SWAP and the thread executing the SWAP is stuck, it is not possible to recover the original node. 
We are also not aware of any lock-free reclamation schemes that handle the SWAP instruction: the hazard pointers and similar schemes cannot be applied as they need to restart reading a pointer.
Recall that restarting a SWAP instruction that reads a pointer might foil the correctness of the data structure.
The lock-free reference counting cannot be applied as it cannot SWAP and increment the reference count of the target node in a single operation. 
The \aoa{} scheme cannot be applied as SWAP cannot be represented in the normalized form. 

Finally, the \freeaccess{} scheme does not support a data structure that uses the SWAP instruction, whereas, it is easy to modify the data structure --- in a completely mechanical way --- so that \freeaccess{} can be applied. 
This can be done by replacing the SWAP instruction with a loop of (boolean) CASes, as illustrated on Listing~\ref{listing-swap}. 
\begin{lstlisting}[label={listing-swap},caption={Emulate SWAP}]
Node *SWAP(atomic<Node*> *addr, Node *reg){
	do{
		Node *cur = addr->load(memory_order_relaxed); 
	}while(!addr->compare_exchange_strong(cur, reg))); 
	return cur; 
}
\end{lstlisting}^^A \label{listing-swap}

We believe this solution is not ideal as it can reduce the progress guarantee of the modified data structure from wait-freedom to lock-freedom (a thread can be delayed an arbitrary number of steps under a worst-cast scheduling) and it might reduces performance. 
Still, as no such lock-free data structure is currently available and no better solution is known, we consider this to be an acceptable solution. 
If SWAP becomes common in the future, it would be interesting to study how it could overcome the difficulty of a stuck thread and whether \freeaccess{} could be modified to handle this usage,
without foiling the wait-freedom guarantee of the data structure. 

Additional instructions that do not satisfy the \freeaccess{} assumption are (i) the CAS version that returns the old value of the field and (ii) the FAA instruction on a field that smartly encodes a counter and a pointer in a single 64-bit word. 
CASes are typically used to modify the shared memory in a synchronous manner, hence the old value of the CAS is typically not important to the data-structure algorithm. 
Still, if a data structure uses the old value returned by a CAS or uses FAA on a pointer field, \freeaccess{} can be applied if the unsupported instruction is replaced by a CAS loop. 

\section{Related Work}\label{sec-related}
In this section, we discuss existing memory-reclamation schemes for non-blocking data structures. 
We consider three aspects of each of these schemes: performance, non-blocking guarantees, and the simplicity of applying them to a given data structure. 
A table summarizing these schemes appear in Appendix~\ref{appedix-related}.  

The first memory reclamation scheme for lock-free data structures was based on reference counting \cite{valois1995lock,mich95}. 
It is the only scheme that can be applied to general code (under the assumption of Section~\ref{subsection-write-only-or-read-only}). 
It offers full lock-freedom support. 
The main problem of the reference-counting scheme is performance: it adds significant overhead and prevents scalability \cite{hart2007performance}. 
The reclamation of cyclic garbage is another limitation.

\citet{harrisll} designed the epoch-based reclamation (EBR) method based on the RCU paradigm. 
The EBR scheme permits threads to have different epoch numbers, but the maximum difference between all threads in the system is at most one. 
Threads reclaim nodes that were retired at least two epochs ago, thus ensuring that all threads finish executing all operations on that epoch. 
EBR is not lock-free, because if a thread does not advance its epoch number, other threads in the system cannot advance the global epoch number, thus preventing reclamation. 
Eventually, the system will run out of memory. 
EBR is easy to apply, but still requires the programmer to insert retire calls, which make the node a candidate for reclamation. 

The hazard-pointers (HP) scheme \cite{HP,mich02} was presented in Section~\ref{subsec-hazardpointers}. 
HP provides full lock-freedom support: threads never wait for other threads to respond. 
However, it often suffers from a high overhead (Section~\ref{sec-measurements}). 
In addition, applying HP is complex; it cannot be applied to standard lock-free data structure such as Harris's linked list \cite{mich02}. 
Instead, it requires changing the data structure, reducing its scalability and performance, and changing its properties (Section~\ref{subsec-hazardpointers}).

The drop the anchor (DTA) scheme \cite{dropta} was designed to improve the performance of the hazard-pointers scheme for linked-list-based data structures.
DTA publishes a single anchor node that implicitly publishes a set of $k$ nodes reachable from the anchor. 
A freezing mechanism is used to recover a delayed thread. 
DTA offers full lock-freedom support. 
It is faster than the HP scheme on long linked lists but suffers from scalability issues on smaller lists \cite{Cohen2015}. 
It is required to re-design a drop-the-anchor scheme for each data structure, and the authors presented such a design only for the linked-list data structure. 

ThreadScan \cite{Alistarh2015a} reduces the overhead of the HP scheme by using OS signals. 
During a reclamation, a signal is sent to all threads in the system; each thread traces its own roots and reports them to the reclamation process. 
The restrictions posed by ThreadScan on the data structures are the same as these posed by the HP scheme. 
ThreadScan cannot be considered fully lock-free as it must wait for all threads to respond before proceeding with the reclamation. 

DEBRA+ \cite{Brown2015} extends the EBR scheme to tolerate threads failures. 
If a thread is delayed and fails to progress its epoch number, other threads naturalize the delayed thread by sending it an OS signal. 
When a naturalized thread wakes up, it must restart the current operation. 
The algorithm never waits for a thread: once a thread is naturalized, other threads can continue executing their operations. 
Nevertheless, DEBRA+ uses the OS signaling mechanism, which acquires locks in the Linux implementation. 
Applying a DEBRA+ scheme to a data structure requires to re-design the data structure and develop a helping mechanism for naturalized threads. 

QSense \cite{Balmau2016} combines the HP and EBR schemes:
When all threads respond, it uses EBR; when a thread fails to respond, it falls back to using HP. 
To avoid the cost of memory barriers at the fast path, it uses rooster processes that the OS scheduler is required to run at fixed intervals. 
There is one rooster process per worker thread and it runs on the same processor core as the thread. 
Its task is to flush, at these intervals, the write buffer of the processor core. 
QSense does not support strict lock freedom as it assumes that threads respond within a fixed time interval
and relies on the lock-freedom of the OS scheduler. 
It poses the same restrictions on the data structure as the HP scheme does. 

Fence eliding techniques for the HP scheme were considered by \citet{Dice2016a}. 
The basic idea is to avoid publishing the hazard pointers during the data-structure operation. 
Only during reclamation is a global synchronization barrier executed, which implies a memory fence of each of the participating threads. 
They offer three such mechanisms. First, they use OS services to execute a global-synchronization barrier. 
Such an OS service does not purely adhere to the lock-freedom model as it assumes that a thread can be interrupted in a fixed amount of time. 
The Linux implementation takes locks to execute this operation. 
A second mechanism relies on a x86-specific hardware global-synchronization barrier; this barrier remains only to support legacy code. 
Such hardware support is not available on other architectures and it might not be supported on future x86 processors. 
A third mechanism is a hardware-assisted hazard-pointers mechanism. 
All these mechanisms pose the same restrictions on the data structure as the HP scheme does. 

The Optimistic Access scheme \cite{Cohen2015} improves the hazard-pointers scheme by replacing publishing hazard pointers, which requires a memory fence, by checking a dirty flag. 
The handling of read instructions in this paper is based on this mechanism. 
If the dirty flag is found to be set, the thread restarts the operation. 
They also concisely define restarting an operation by using the normalized form of \citet{lf2wf}. 
The Optimistic Access scheme poses the same restrictions on the data structure as the HP scheme does. 

\ignore{The Automatic Optimistic Access scheme \cite{Cohen2015c} attempted to simplify the task of 
applying memory reclamation support to a given data structure. 
It is based on the Optimistic Access scheme for handling reads efficiently. 
During reclamation, it employs a lock-free mark and sweep algorithm, adapted from the garbage collection literature to be fully lock-free. 
This scheme is simple to apply and and does not require the programmer to insert retire calls. 
However, it requires the data structure to be presented in a normalized form so that restarts are well-defined. 
While most lock-free algorithms in the literature can be modified to be in the normalized form, 
it is unknown whether future algorithms would also follow the normalized representation. 
It does not support atomic instructions other than CAS, such as fetch-and-add. 
Modifications to the data structure are inserted manually by the programmer. }

Hazard Eras~\cite{ramalhete2017brief} and Interval-Based Reclamation~\cite{wen2018interval} use a combination of epoch-based reclamation and hazard pointers schemes. 
Each thread publishes the current epoch it observes. 
These schemes do not reclaim any node that was alive during a published epoch, but can reclaim nodes removed before such an epoch or allocated after such an epoch. 
Both these schemes pose the same restrictions on the data structure as the HP scheme does. 

Alistarh et al. \cite{Alistarh2017} presented ForkScan, a garbage collection scheme that uses programmer hints to achieve scalability. 
These hints are checked by the garbage-collection system, hence the programmer does not have to restrict itself in the data structure design. 
Similar to on-the-fly garbage-collection algorithms, ForkScan relies heavily on OS services, both to interrupt threads in order to gather roots and to 
provide a snapshot of the heap, hence it cannot be considered fully lock-free. 

The \freeaccess{} scheme relies on the ability to restart read-only periods. 
This technique is used by Java's StampLock~\cite{stampedlock} for synchronizing optimistic readers and was later used by Layout Lock~\cite{Cohen2017layout} for synchronizing structural changes to a data structure. 

\section{Conclusion}\label{sec-conclusion}
In this paper, we presents \gaoa{}, a lock-free memory-reclamation scheme that is applicable to general data-structure code. 
The \freeaccess{} scheme does not require the data structure to be presented in a special normalized-form or support a special restart mechanism.
Furthermore, applying \freeaccess{} is easy because code modifications are done automatically by a compiler plug-in.
This enables a separation of concerns: Memory reclamation does not prevent designs of novel lock free data structures and does not prevent a practitioner from applying these data structures in real applications. 
The overhead of the \freeaccess{} scheme is surprisingly low: it always outperforms the automatic optimistic access scheme and in many cases also the hazard-pointers scheme (with and without fence elision). 

\bibliography{FreeAccess,LockFreeMM,raoa,book,references}
\appendix
\newpage
\section{Existing Memory Reclamation Schemes for Lock-Free Data Structures}\label{appedix-related}

\begin{table}[h]
	\centering
	\footnotesize
	\caption{Existing lock-free memory reclamation schemes}
	\label{my-label}
	\begin{tabular}{|l|l|l|l|l|}
		\hline
		Scheme                                                             & Based on                                                         & Performance                                               & Simplicity                                                                                                & \begin{tabular}[c]{@{}l@{}}Lock Freedom\\ Dependences if exist\end{tabular}                                            \\ \hline
		RC [V95,MS95]\ignore{\cite{valois1995lock,mich95}}                                  &                                                                  & Very slow                                                 & \begin{tabular}[c]{@{}l@{}}Automatic \\ (Section \ref{subsection-write-only-or-read-only})\end{tabular} & Yes                                                                                                                    \\ \hline
		EBR [H01]\ignore{\cite{harrisll}}                                              & RCU                                                              & Fast                                                      & Simple                                                                                                    & No                                                                                                                     \\ \hline
		HP [M02a,M04]\ignore{\cite{mich02,HP}}                                              & RC                                                               & Slow                                                      & Complex                                                                                                   & Yes                                                                                                                    \\ \hline
		DTA [BKP13]\ignore{\cite{dropta}}                                                & HP                                                               & \begin{tabular}[c]{@{}l@{}}Fast \\ (depends)\end{tabular} & Very complex                                                                                              & Yes                                                                                                                    \\ \hline
		StackTrace [ALMS15]\ignore{\cite{Alistarh2015a}}                                   & HP                                                               & Fast                                                      & Complex (HP)                                                                                              & \begin{tabular}[c]{@{}l@{}}Partial. Wait for threads to respond. \\ Depends on OS signals.\end{tabular}                \\ \hline
		OA [CP15b]\ignore{\cite{Cohen2015}}                                              & HP                                                               & Fast                                                      & Complex (HP)                                                                                              & Yes                                                                                                                    \\ \hline
		DEBRA+ [B15]\ignore{\cite{Brown2015}}                                          & EBR                                                              & Fast                                                      & Complex                                                                                                   & \begin{tabular}[c]{@{}l@{}}Partial. \\ Depends on OS signals and \\ stack unwinding (longjmp).\end{tabular}            \\ \hline
		AOA [CP15a]\ignore{\cite{Cohen2015c}}                                            & \begin{tabular}[c]{@{}l@{}}OA\\ M\&S GC\end{tabular}             & Fast                                                      & \begin{tabular}[c]{@{}l@{}}Simple\\ Normalized form\end{tabular}                                          & Yes                                                                                                                    \\ \hline
		QSense [BGHZ16]\ignore{\cite{Balmau2016}}                                         & HP, EBR                                                          & Fast                                                      & Complex (HP)                                                                                              & \begin{tabular}[c]{@{}l@{}}Partial. \\ Assumes threads interact in fixed time.\\ Depends on OS scheduler.\end{tabular}    \\ \hline
		FenceElision [DHK16]\ignore{\cite{Dice2016a}}                                    & HP                                                               & Fast                                                      & Complex (HP)                                                                                              & \begin{tabular}[c]{@{}l@{}}Partial. Wait for threads to respond.\\ Depends on OS interrupts.\end{tabular}              \\ \hline
		HazardEras [RC17] & HP, EBR & Fast & Complex (HP) & Yes \\ \hline
		IBR [WICBS18] & HP, EBR & Fast & Complex (HP) & Yes \\ \hline
		ForkScan [ALMS17]\ignore{\cite{Alistarh2017}}                                     & \begin{tabular}[c]{@{}l@{}}StackTrace\\ OnTheFly GC\end{tabular} & Fast                                                      & \begin{tabular}[c]{@{}l@{}}Simple\\ Programmer hints\end{tabular}                                         & \begin{tabular}[c]{@{}l@{}}Partial. Wait for threads to respond. \\ Depends on OS signals. Copy-on-write.\end{tabular} \\ \hline
		\begin{tabular}[c]{@{}l@{}}FreeAccess \\ (this paper)\end{tabular} & AOA                                                              & Fast                                                      & \begin{tabular}[c]{@{}l@{}}Automatic \\ (Section \ref{subsection-write-only-or-read-only})\end{tabular}                                            & Yes                                                                                                                    \\ \hline
	\end{tabular}
\end{table}

\section{Hardness of Applying the Hazard-Pointers scheme}\label{appendix-hp-bad}
%
%
%
The Harris's linked list is incompatible with the hazard-pointers scheme (as reported by the original hazard pointers paper \cite{mich02}). 
The reason is similar to the one described in Section~\ref{subsec-hazardpointers}, but the details are somewhat different and provided here for completeness. 
Consider a snippet of a linked list with four nodes $\boxed{1} \Rightarrow \boxed{2} \Rightarrow \boxed{3} \Rightarrow \boxed{4}$ and suppose that both \boxed{2} and \boxed{3} were marked as removed from the list, but were not yet disconnected from the list (so \boxed{1}'s next pointer still points to \boxed{2}). 
In such a case, it clearly make sense to remove both nodes (\boxed{2} and \boxed{3}) by using a single CAS instruction that modifies \boxed{1}'s next pointer to \boxed{4}. 
This optimization is incorporated in Harris's linked-list algorithm: when a marked node is found, the algorithm attempts to find additional sequentially marked nodes and to disconnect all these nodes together. 

Next, suppose that no memory reclamation is used, and that a thread $T$ currently traverses node \boxed{2}. 
As \boxed{2} is marked, $T$ would read \boxed{2}'s next pointer (\boxed{3}) and find that it is also marked. 
Then it would read \boxed{3}'s next pointer (presumably \boxed{4}) and find (by reading \boxed{4}'s content) it is not marked. 
Finally, $T$ would attempt to CAS \boxed{1}'s next pointer from \boxed{2} to \boxed{4}. 

This technique does not work when the hazard-pointer scheme is used. 
When $T$ traverses \boxed{2}, it only protects \boxed{2} via a hazard pointer. 
A concurrent thread might have disconnected both \boxed{2} and \boxed{3} from the list and retire both. 
Since only \boxed{2} is guarded by a hazard pointer, \boxed{3} would be reclaimed and its content becomes arbitrary. 
When $T$ resumes its operation, it would read \boxed{3}'s next pointer, but obtain an arbitrary address (instead of \boxed{4}). 
Then, it would crash when checking, which is a read from an arbitrary address, whether the node is marked.


\end{document}